\documentclass[preprint]{elsarticle}

\usepackage{hyperref}
\usepackage{booktabs,dcolumn}
\newcolumntype{d}[1]{D..{#1}}

\usepackage{booktabs}
\usepackage{amsfonts}

\usepackage{array}
\usepackage{graphicx}
\usepackage{colortbl}
\usepackage{multirow}
\usepackage{pdflscape}
\usepackage{amsmath}
\newcommand{\tabitem}{\llap{\footnotesize\textbullet}~}

\journal{Applied Energy}

\biboptions{numbers,sort&compress}

\usepackage{amsmath}

\begin{document}

\begin{frontmatter}

\title{Ancillary Services Acquisition Model: considering market interactions in policy design}
\author[mymainaddress,mysecondaryaddress]{Samuel Glismann\fnref{myfootnote}}
\fntext[myfootnote]{M.Eng., PhD candidate, Lead in System Operations}

\ead{samuel.glismann@tennet.eu}

\address[mymainaddress]{Europa-Universität Flensburg, Auf dem Campus 1, 24943 Flensburg, Germany}
\address[mysecondaryaddress]{TenneT TSO B.V., Utrechtseweg 310, 6812 AR Arnhem, The Netherlands}

\begin{abstract}
A rapidly changing electricity sector requires adjusted and new ancillary services, which enable the secure and reliable operation of the electricity system. However, assessments and policy advice regarding ancillary services and market design lack methods to evaluate the complex interaction of markets and services. Therefore, this paper contributes an open-source agent-based model to test design options for ancillary services and electricity markets. The \textit{Ancillary Services Acquisition Model} (ASAM) combines the agent-based modeling framework MESA with the toolbox Python for Power System Analysis (PyPSA). The model provides various design parameters per market and agent-specific strategies as well as detailed clearing algorithms for the day-ahead market, intra-day continuous trading, redispatch, and imbalances. Moreover, ASAM includes numerous policy performance indicators, including a novel price mark-up indicator and novel redispatch performance indicators. A stylized simulation scenario verified and validated the model and addressed a redispatch design question. The results displayed the following implications from order types in redispatch markets with multi-period all-or-none design: (1) The order design provides few risks for market parties, as orders are fully cleared. (2) Large orders may lead to dispatch ramps before and after the delivery period and may cause “ramp-risk” mark-ups as well as additional trading of imbalances on intra-day. (3) All-or-none design in a liquid situation leads to the over-procurement of redispatch by the grid operator, as orders cannot be partially activated. Moreover, it is likely that the grid operator induces imbalances to the system by “incomplete” redispatch activation (i.e. upward and downward redispatch volumes are not equal).

\end{abstract}


\end{frontmatter}

\section{Introduction}
The transition of electricity systems briskly materializes with phenomena such as renewable energy sources \cite{EurObservER2018}, demand-side response \cite{JointResearchCentre2016}, and new business models \cite{Amelang2019}. 
In this context, policymakers need to adapt existing ancillary services\footnote{The "ancillary services" definition from \cite{Glismann2017} is used: “\textit{all services that system users provide to the system operator and the network operator for the operation of the transmission system and the distribution system}”.}, to cope with new physical challenges (e.g. changing ancillary service demand) and new market situations (e.g. changing ancillary service supply). Examples of ancillary services in Europe are frequency restoration reserves \cite{ENTSO-E2020}, redispatch services \cite{Hirth2018}, and reactive power services \cite{Anaya2020}. An inadequate design of ancillary services can restrain operators from effectively providing access to grid and system users, and eventually lead to interruptions of the electricity supply \cite{Nobel2016}.
\\
Various authors have published design frameworks for electricity markets and ancillary services \cite{Doorman2011,Abbasy2012,VanderVeen2012, Iychettira2017,Poplavskaya2019}. However, the interactions of markets and ancillary services have remained particular challenges for assessments of policy performance; the model complexity increases as well as the modeling and simulation effort of researchers \cite{Tesfatsion2018}. 
 
Therefore, this paper introduces a generic, state-of-the-art model to facilitate market and ancillary services design studies. The contributions to the research community are as follows:
\begin{itemize}
     \item An open-source and comprehensive agent-based model covering short-term markets and ancillary services with configurable design variables in the European context, that is day-ahead market (DAM), intra-day market (IDM), redispatch mechanism (RDM), balancing energy market (BEM), and imbalance mechanism (IBM)\footnote{The specific redispatch and imbalance design determines whether it is referred to as a mechanism or as a market.}.
     \item Market clearing algorithms, that allow for a detailed simulation of order types (i.e. limit orders, all-or-none orders, and multi-period block orders). This is a novelty for redispatch market models, which generally highly simplify the procurement process.
     \item Multiple novel policy performance indicators regarding market interactions, price mark-ups of market parties related to various risks and opportunities, and the grid operators' redispatch performance (i.e. over-procurement, under-procurement, redispatch induced imbalance).  
     \item The open-source combination of the agent-based modeling framework MESA with the toolbox Python for Power System Analysis (PyPSA).
\end{itemize}
\leavevmode\newline
The remainder of this paper is structured as follows: Section \ref{sec: material and methods} introduces related literature and methods as well as the choice for an agent-based modeling method. Section \ref{sec: theory/calculation} provides a detailed description of the aim, structure and features of the ancillary services acquisition model (ASAM). Section \ref{sec: results} explores and verifies simulation results of a stylized scenario, followed by a discussion and the conclusion in sections \ref{sec: discussion} and \ref{sec: conclusion}, respectively.
\section{Related literature}
\label{sec: material and methods}
 \subsection{Fundamental electricity market design in Europe}
 In Europe, market parties have the general freedom to schedule dispatch of their generator, consumption and storage assets for their own benefit (i.e. "self-dispatch") \cite{Nobel2016,Hirth2018}. A "zonal-pricing" approach (as oppose to "nodal-pricing") allows market parties to trade with any other market party within a bidding zone, without consideration of grid constraints. Trading across bidding zones is subject to cross-border transfer capacity and organized in market-coupling processes \cite{EuropeanCommission2015}. Trading starts years prior to delivery with the forward markets and continues until "real-time" and even ex-post trading. On the day before delivery, a single double-sided auction\footnote{Double-sided implies multiple buyers and sellers, whereas single-sided auctions have either one buyer or one seller.} constitutes the DAM, followed by continuous trading of the IDM\footnote{IDM actions will be introduced to continuous trading, in line with EU regulation \cite{EuropeanCommission2015}.}. In parallel with the wholesale market processes, transmission system operators (TSOs) and Distribution System operators (DSOs) acquire ancillary services. For example: the balancing capacity market (BCM) is typically organised before DAM, redispatch (mainly) takes  place in parallel with IDM \cite{EuropeanCommission2015}, and the BEM is a single-sided real-time auction \cite{TheEuropeanCommission2017}. Ex-post, the IBM financially settles the imbalance of market parties (i.e. the delta between the energy that has been injected or withdrawn from the grid and the energy that has been traded per imbalance settlement period - ISP) \cite{TheEuropeanCommission2017, Nobel2016, TenneTTSOB.V.2019}.          
The reader can refer to \cite{Brijs2017} for a more detailed review of these sequential markets in Europe. 
\subsection{Perspectives of ancillary services research}
Existing studies and models regarding ancillary services may be distinguished by the following perspectives:
\begin{itemize}
    \item Economic self-optimization of certain market participants or technologies (e.g. wind power plants \cite{Hosseini2020}, storage systems \cite{Merten2020}, demand facilities \cite{Herre2020}) in various electricity markets and ancillary service processes.
    \item Physical performance of the power system and technological feasibility (e.g. impact of the share of renewable energy sources on inertia \cite{Mehigan2020,Johnson2019})
    \item Policy advice for ancillary services and market design (e.g. capacity remuneration versus energy only market \cite{Keles2016}, bidding zone sizes \cite{Ambrosius2020} and congestion management \cite{Hladik2020} in Europe, pricing mechanism for balancing energy and imbalance \cite{Nobel2016}, and procurement strategies for system operators \cite{Greve2018})
\end{itemize}
The aim of the present paper relates to the policy advice perspective.

\subsection{Agent-based modeling in electricity system research}
Agent-based models are today one simulation approach next to equilibrium models and optimization models (see \cite{Ringkjob2018} for a model review). Sources \cite{Weidlich2008} and \cite{Tesfatsion2018} have substantially contributed to the incorporation of valid agent-based analyses into energy system research. Fields of application include the investigation of bidding strategies in electricity markets \cite{VanderVeen2012a, Wehinger2013,Rashedi2016,Rashidizadeh-Kermani2018}, decision-making in grid expansion investments \cite{Blijswijk2017}, benefit distributions as well as the effectiveness of policies for renewable energies expansion \cite{Reeg2013, Nunez-Jimenez2020, Ramshani2020}, and the interaction of different electricity markets and ancillary services \cite{Weidlich2008,Petitet2019,Poplavskaya2020}. 
\\
Multiple characteristics of agent-based models support their suitability for policy design testing: (1) The "bounded rationality" of agents enable robustness tests of policy options with "imperfect" behaviour \cite[p. ~104]{Ostrom2005}. (2) agent-based models can be combined with classical optimization methods (e.g. linear-programming) as well as with self-learning methods. The authors of \cite{Klein2019} call this characteristic "models within models". (3) agent-based models can also provide bottom-up exploration and communication for policymakers, by showing emergent system behaviour as a result selected assumptions on agent level \cite{Klein2019}.
\subsection{Availability of open-source models}  
In recent years, various open-source and open-access energy system models (e.g. Oemof) and power system analyses tools (e.g. PyPSA) became available (see open energy models initiative \cite{openmod}). PyPSA is a renown python-based toolbox, that includes linear least-cost optimization of power plant and storage dispatch within network constraints \cite{pypsa_article, pypsa_docu}.
\\
Still, comprehensive agent-based models like AMIRIS, SiStEM and Enertile (previsously named PowerACE) are not openly available for researchers. The open-source Java-based AMES is an exception. The model is developed as a "test bed" for policies regarding United States (U.S.) market design \cite{Battula2020}. 
\\
The literature review showed that, currently, an open-source agent-based model does not exist for design tests of electricity markets and ancillary services in the European Union (EU) context.  
\section{Model description}
\label{sec: theory/calculation}
\subsection{Aim, scope, and main assumptions}
\label{sec:Aim, scope and main assumptions}
In order to evaluate the consequences of various ancillary service acquisition\footnote{The term acquisition is chosen instead of procurement, as ancillary service designs may be obligatory and non-remunerated \cite{Glismann2017}} designs, including potential interactions with other acquisition processes and markets, the model aims to explore both agent behavior and emergent system behavior. The model supports assessments of policy implications by simulating designs which may not (yet) be available in the real-world. Hence, the model may produce out-of-sample results \cite{Windrum2007}. Given these objectives, the model is called ancillary service acquisition model (ASAM).

The main agents of the model are "market parties", "grid \& system operators" and "market operators". These agents may be further specified in sub-agents, if needed. Regulators are not considered in this operative model.  
Grid \& system operators are agents with the responsibilities of TSOs and DSOs (typical EU roles) or independent system operators and transmission/distribution utilities (typical U.S. roles). Grid \& system operators are considered as non-competitive, monopolistic, and highly regulated agents. 
Market operators are agents that facilitate wholesale market processes such as receiving orders, matching orders and settling transactions. For the purpose of this model, market operator agents additionally operate ancillary service acquisition processes, even though these processes are often executed by grid \& system operators or grid \& system operator owned entities (e.g. Joint Allocation Office). 

ASAM has the following characterizing features:
\begin{enumerate}
\item A model structure that supports the implementation of various market processes from forward markets up to real-time markets with various designs. 
\item Market party agents own assets (generation, consumption and storage facilities) with configurable (inter-temporal) constraints.
\item Market party agents may have individual heuristic strategies to act on market or acquisition processes. It is assumed that agents act with bounded rationality, i.e. they "\textit{are goal oriented and try to be rational but face cognitive limits}” \cite[p. ~104]{Ostrom2005}. Their strategies per market concern quantity, price and the timing (relative to the delivery period) of orders. 
\item Rules for markets and ancillary service acquisition processes are configurable. The generic design variables per market or acquisition process are acquisition method, pricing method, order types, gate-opening time, gate closure time, and provider accreditation (i.e. pre-qualification for specific ancillary services. See also \cite{Glismann2017}).
\item A reporting class allows for the implementation of performance indicators that measure the impact of design options. 
\item ASAM's generic modules for the simulation of ancillary services acquisition and markets are well-suited to be combined with detailed models of physical system behaviour (e.g. load-flow models and frequency models). Currently, ASAM is already linked to PyPSA.  
\item Simulation complexity is adjustable with options for using exogenous input data regarding ancillary services demand (e.g. redispatch demand) instead of endogenous demand data. Moreover, (residual) load, generator outages, and forecast errors may be provided as exogenous time series.
\item ASAM is open-source and open access to ensure scientific integrity and further development by the science community.
\end{enumerate}

\begin{figure}[th]
\centering
\includegraphics[width=1\textwidth]{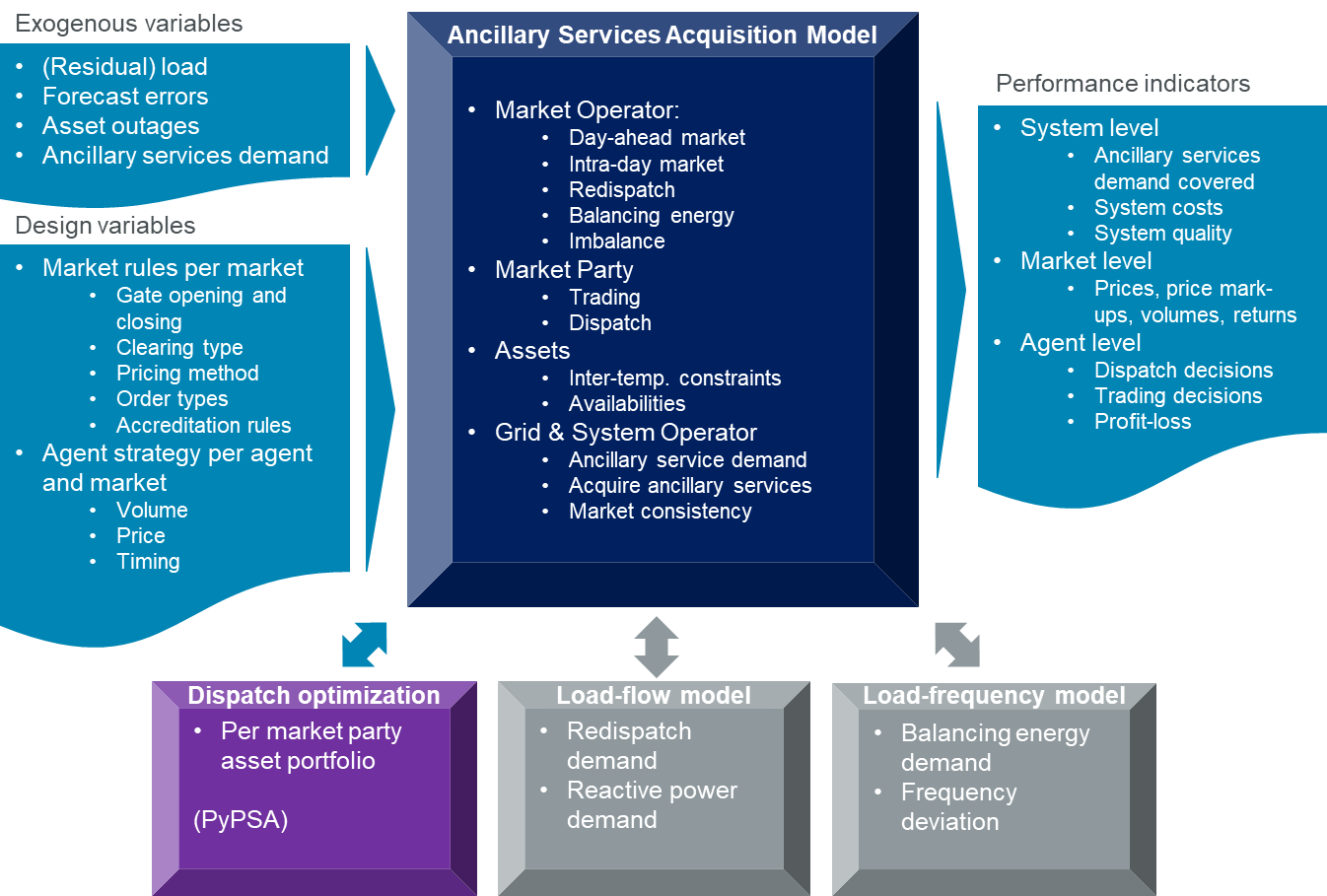}
\caption{ASAM overview and link to other models (only PyPSA is currently implemented).}
\label{fig:ASAM overview}
\end{figure}
The model incorporates the following (Europe-centric) market design assumptions and theory: 
\begin{enumerate}
\item ISP of 15 minutes (see \cite{EuropeanUnion2019}). This is also the step-size of the agent-based simulation.
\item Market parties are allowed to conduct self-dispatch and bilateral trading (see \cite{Nobel2016}). 
\item  The DAM is organised via a single, sealed auction with a market time unit (MTU) of one hour\footnote{The MTU of the day-ahead market will be changed to 15 minutes in the coming years \cite{EuropeanUnion2019}} (see \cite{VonSelasinsky2014}). 
\item The IDM is implemented as intra-day continuous trading  with continuous double-sided auctions and with an open order book. The MTU is assumed to be 15 minutes \footnote{This is a simplification, because in some countries also intra-day auctions exist and trading with hourly MTU is available.} (see \cite{VonSelasinsky2014}).
\item Zonal pricing is assumed for electricity markets instead of nodal pricing. Cross-zonal capacity in DAM and IDM is allocated implicitly via market coupling processes (see \cite{Hirth2018}).\footnote{Currently, ASAM has only a one-zone market implemented} 
\item Redispatch service acquisition is considered here as a "pro-active" process (i.e. before an anticipated congestions becomes real) with discrete moments of redispatch (see \cite{Hirth2018}). 
\item Balancing energy orders are activated "reactively" (i.e. after an area control error occurs) in response to the load-frequency controller in real-time (see \cite{TenneTTSOB.V.2019}).
\item Imbalance mechanisms may have a design incentivizing deliberate market imbalances to improve the balance control situation (i.e. passive balancing) \cite{Nobel2016}. Other imbalance mechanisms may have a design to primarily redistribute balancing energy cost to market parties and to penalize any market imbalance (see \cite{TenneTTSOB.V.2019,Nobel2016}).
\end{enumerate}

The following markets and acquisition processes are currently implemented in ASAM: DAM, IDM, RDM, BEM, and IBM. However, forward markets, bilateral over-the-counter trade\footnote{It is assumed that all intra-day trade takes place on continuous trading platforms}, balancing capacity auctions, and other ancillary services can be modeled in future, using ASAM.

The process of retail contracts and transactions between electricity suppliers and consumers is not an explicit part of the model. Additionally, adjacent markets such as fuel markets and CO$_2$ markets are out of scope. 

the characteristics and dispatch optimization of generators and loads is based on \cite{pypsa_article}.

Figure \ref{fig:ASAM overview} displays an overview of the input variables, main functionalities and model output. Figure \ref{fig:ASAM overview} additionally shows the interface with other models that simulate the physical aspects of the electricity system. 
 \subsection{Performance indicator price mark-ups}
Model users can add specific performance indicators to the report class of ASAM in order to test design options. A distinguishing feature of ASAM is the implemented "price mark-up" performance indicator.
“\textit{The efficiency of markets can also be assessed by the presence, or absence, of perverse incentives to those involved. Such incentives may result in ineffective or inefficient behaviour and/or inefficient mark-ups in prices}” \cite[pp. ~11]{Nobel2016}. However,
empirical methods as well as analytical methods regarding the efficiency of markets additionally require benchmark costs to evaluate these inefficient mark-ups. When only assuming fixed cost for all studied ancillary service designs, the study would fail to exhibit the design-driven costs, which are not related to strategic mark-ups but stem from risk and opportunity costs for market parties. Hence, a price mark-up analysis can, bottom-up, reveal the impact of market designs on costs for market parties and thus show the affected market efficiency.

\begin{figure}[th]
\centering
\includegraphics[width=0.6\textwidth]{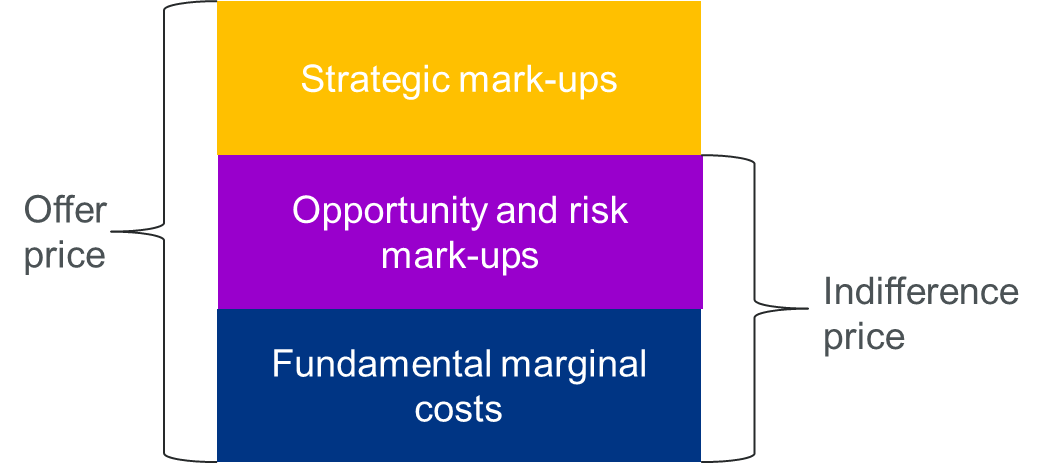}

\caption{Concept of price mark-up model}
\label{fig:mark-up model}
\end{figure}

ASAM distinguishes marginal costs that are fundamental to service delivery\footnote{These costs are rather independent from market or ancillary services design} as well as two types of mark-ups: (1) opportunity and risk mark-ups from the market design, and (2) strategic mark-ups stemming from market power (see figure \ref{fig:mark-up model}).

 As forgone opportunities may also be considered to be risks, equation \ref{eqn: general markup} provides a simple generic form to determine risk and opportunity mark-ups. Market parties choose methods to determine expected risk prices and quantities, depending on their risk appetite and beliefs. Subsequently, the mark-up is added to the order price.
\begin{equation}
RiskMarkup = \frac{(exp.RiskPrice * exp.RiskQuantity)}{OfferQuantity}
\label{eqn: general markup}
\end{equation}

For an example of fundamental costs and opportunity mark-up, the reader can refer to the guidance proposal for redispatch cost remuneration in Germany \cite{bdew2018}. \cite{VonSelasinsky2014} uses the term "indifference price", which is the offer price, including risk mark-ups, where the market party would be indifferent about being matched or not. Hence, any prices beyond this indifference price include strategic mark-ups (see also \cite{ockerPhD} and \cite{BMWi2020} for strategic behavior). 

\subsection {Model structure}
\begin{figure}[th]
\centering
\includegraphics[width=1\textwidth]{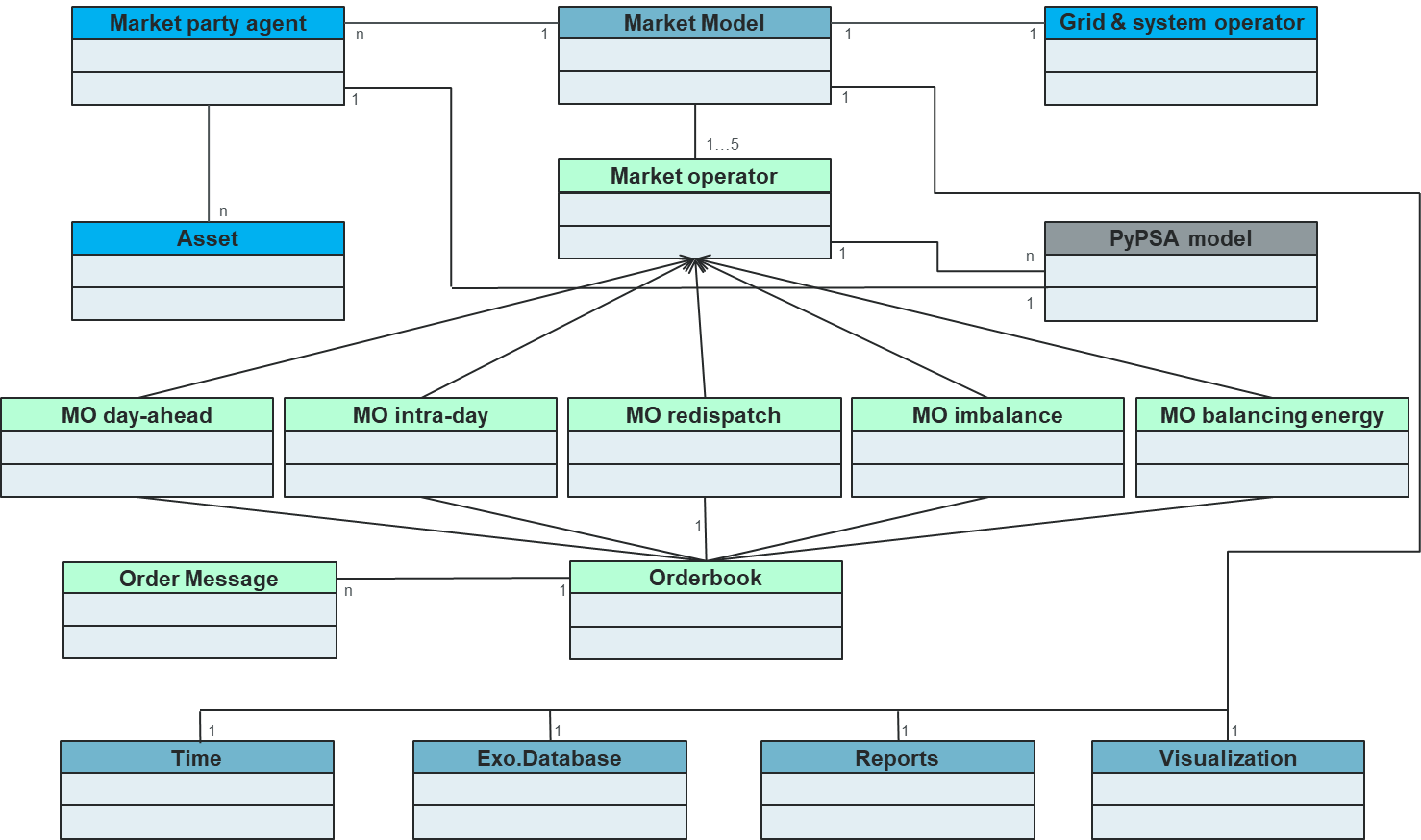}

\caption{Overview of classes in ASAM.}
\label{fig:classes}
\end{figure}

\paragraph{Libraries and classes} ASAM is implemented in Python 3.5.6. It uses various Python libraries. It uses classes of the agent-based modeling framework MESA \cite{MESAweb}, that is model class, agent class, scheduler class. Instances of PyPSA are applied for the dispatch optimization of market parties and for day-ahead and redispatch market clearing. The classes are depicted in Figure \ref{fig:classes}. Methods and attributes per ASAM class are listed on GitHub (see \cite{AsamGithub}). 

\paragraph{The model concept of time} As is common in agent-based simulations, time progresses in discrete steps. During each step, agents take actions, given the stage of the simulation. One simulation step corresponds to one ISP of 15 minutes. The time is expressed in tuples of day ($\in \mathbb{N}$) and ISP of the day ($\in \mathbb{N}\{1,…,96\}$). 
Actions of agents related to short-term markets do not only consider the current simulation step but also all simulation steps up to the last delivery period of the latest day-ahead auction (i.e. inter-temporal agent strategies).

\begin{figure}[!htb]
\centering
\includegraphics[width=1\textwidth]{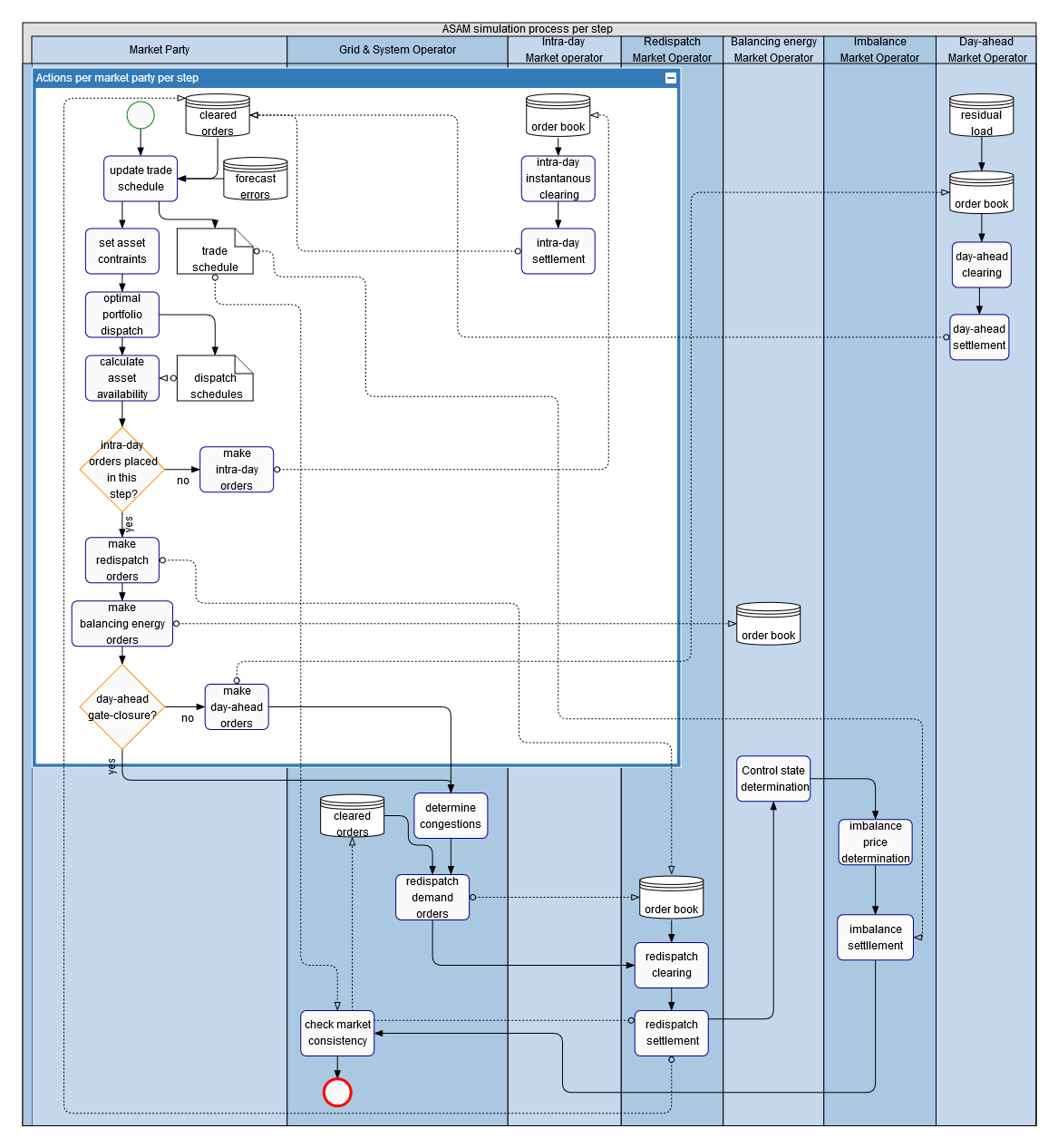}
\caption{Process of one simulation step in ASAM.}
\label{fig:process asam}
\end{figure}
A process overview of one simulation step is illustrated in Figure \ref{fig:process asam}. The following sections explain the process steps of markets and agents.

\subsection {Market and ancillary services implementation}
\paragraph{Day-ahead market single auction} The clearing method of the DAM uses a linear optimal power flow of PyPSA\footnote{ PyPSA version 0.13.1. is used. For a detailed description, the reader is referred to \cite{pypsa_article} and the \cite{pypsa_docu}.} with a one-node grid\footnote{Cross-zonal market coupling may be implemented in future. For an extension, each bidding zone needs to be represented by a node.}. Moreover, exogenous time series of generators’ unavailability are taken into account. A single load represents the buy orders of the market, which corresponds to the exogenous residual load time series. Hence, the DAM is implemented with a single-sided auction instead of the real-world double-sided auction.

\begin{figure}[th]
\centering
\includegraphics[width=1\textwidth]{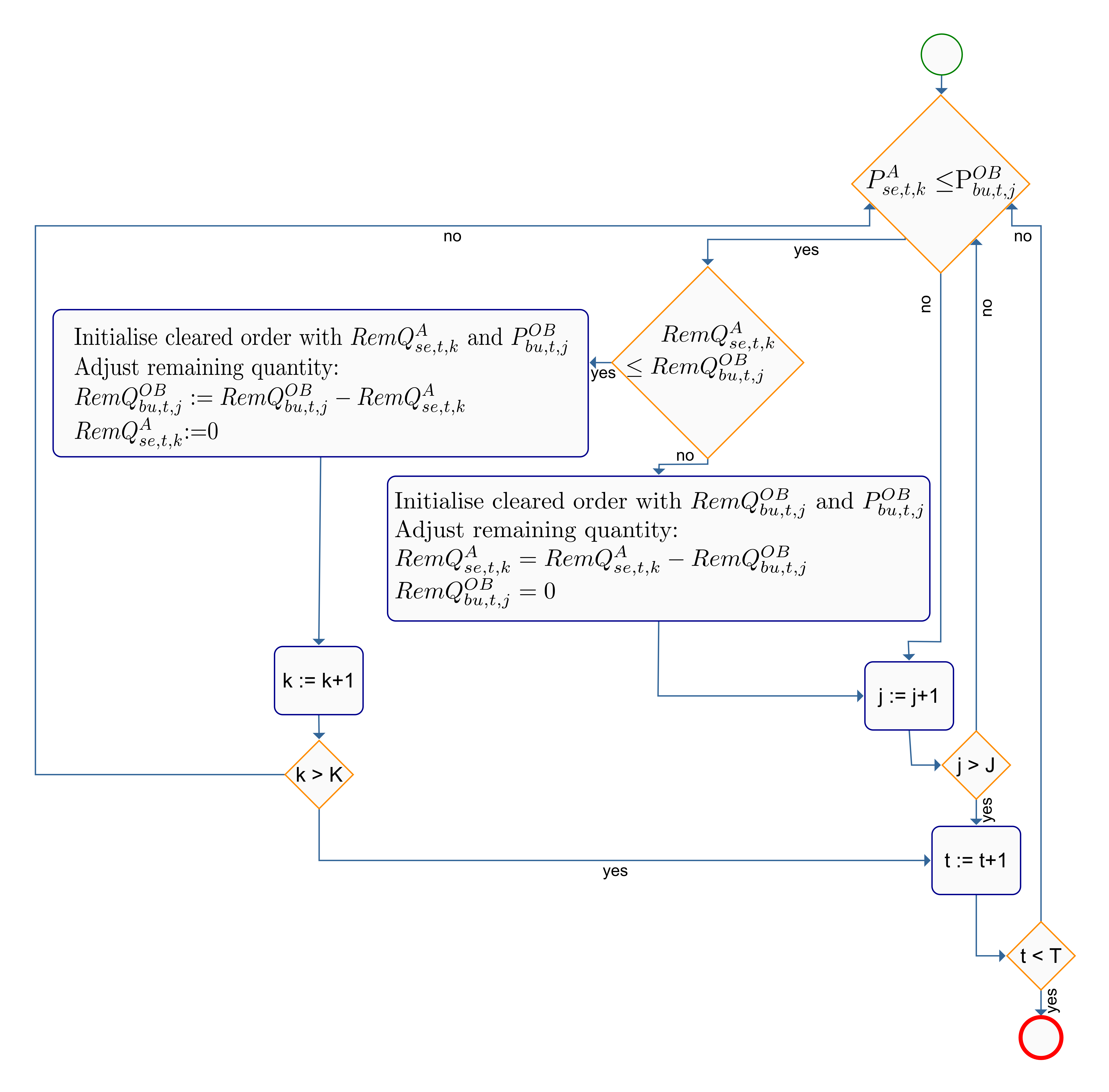}

\caption{Intra-day matching process with the example of sell orders placed by agent A.}
\label{fig:IDalgo}
\end{figure}

\paragraph{Continuous intra-day market} As shown in the process diagram, the continuous intra-day market is instantaneously cleared when an order message is added \cite{VonSelasinsky2014}. Therefore, the intra-day order book may be cleared multiple times per simulation step (i.e. each time an agent adds intra-day orders). The clearing method has three steps:
	
\begin{enumerate}
\item Filter all relevant orders of the order book for matching the new orders. This step results in four sorted sets.
    \begin{enumerate}
    \item 	$P_{se,t,l}^A$ and $P_{bu,t,m}^A$ are sets of order prices of agent A triggering the clearing which have the direction sell ($se$) and buy ($bu$), respectively, and the delivery period $t$. Whereby $k \in\{1,…K\}$  and $l \in\{1,…L\}$ are the indexes of the order sets. $P_{se,t,k}^A$ is sorted by ascending order price. $P_{bu,t,l}^A$ is sorted by descending order prices. 
    \item 	$P_{se,t,i}^{OB}$ and $P_{bu,t,j}^{OB}$ are sets of order prices in the order book OB which have the direction sell ($se$) and buy ($be$), respectively, and the delivery period t. Whereby $i \in\{1,…I\}$  and $j \in\{1,…J\}$ are the indexes of the order sets. $P_{se,t,i}^{OB}$ is sorted by ascending order price and by ascending init\_time. $P_{bu,t,j}^{OB}$ is sorted by descending order prices price and by ascending init\_time. 
    \item Analogously to the notation above the sets $RemQ_{se,t,k}^A$, $RemQ_{bu,t,l}^A$,  $RemQ_{se,t,i}^{OB}$, and $RemQ_{bu,t,j}^{OB}$ are defined, which refer to the remaining quantity of the respective order. 
    \end{enumerate}
\item  Match of new orders with a clearing price of the ‘older’ order (i.e. with the lower init\_time). The matching process of intra-day orders is illustrated in Figure \ref{fig:IDalgo}. Note that the clearing method only allows for order duration\_time of one ISP.
\item  Remove matched orders and adjust quantities of partially matched orders. Keep unmatched limit orders and remove unmatched market orders, as the latter has an immediate-or-cancel constraint. 
\end{enumerate}

\paragraph{Redispatch market implementation} Redispatch is applied by grid operators to shift scheduled load and consumption from one location to another in order to relieve grid congestions \cite{Hirth2018}. ASAM includes a simplified redispatch market scoring algorithm \footnote{Although the terms "redispatch market" and "redispatch orders" are used, the model functionality may also be used for highly regulated redispatch mechanisms}. As a starting point, it has to be noted that the physical grid is currently not modeled in ASAM. The model considers physical characteristics of power plants and residual load. An (approximate) redispatch demand is, therefore, to be derived from congestion and then provided as exogenous data to the model. The grid operator agent places this demand in the redispatch order book. Redispatch demand is defined as a MW value from an area (downward) to an area (upward). Given this exogenous redispatch demand approach, market parties cannot influence the redispatch demand with their behavior.

The redispatch market operator class uses PyPSA to find a redispatch solution. Each redispatch area that is defined in the redispatch demand is initiated as a node in a power system without connections to other nodes. The redispatch demand per area is implemented as a load, and redispatch supply orders are implemented as generators, using generator parameters to emulate order attributes, for example p\_min\_pu is used to define whether an order can be partially called (i.e. limit order) or only fully called (i.e. all-or-none).\footnote{ As an example: The Dutch redispatch design has currently two products: Reserve Other Purposes with all-or-none orders (see \cite{TennetROP}) and a shared DSO-TSO product called IDCONS with limit orders (see \cite{gopacs2019}).}

The optimal power flow function of PyPSA with its inter-temporal constraints calculates which orders are cleared (and to what extent). Slack generators in all areas capture under-procurement and over-procurement of the redispatch market.

This novel redispatch clearing algorithm allows for simulating two market design features that are usually strongly simplified due to complexity: 
\begin{itemize}
\item One customized constraint added to the solver allows for the clearing of multi-MTU block orders (see \cite{TennetROP}).
\item Another customized constraint is a threshold for the "equilibrium constraint" as described in \cite{Hirth2018}: “\textit{In order to avoid (large) imbalances as a consequence of redispatch, network operators aim to apply downward redispatch volume (MWh) and upward redispatch volume per imbalance settlement period (ISP) in equilibrium}” (p.17). The threshold defines the maximum delta between the cleared quantities per MTU in upward direction and in downward direction.
\end{itemize}
The mathematical formulation of the algorithm is available in \ref{apx: redisaptch algo}.

\paragraph{Balancing energy market implementation} The balancing energy market is currently implemented with a high degree of simplifications. As of yet, the market operator has neither a clearing method nor a settlement method. Exogenous balancing energy activation data or a (physical) balancing model would be required for a clearing algorithm. However, the implemented order book allows for the analyzing of received orders, which may provide insight into market interactions. Moreover, a method is available to allow the market operator to determine a balancing control state per simulation step, based on an exogenous probability distribution. In the absence of activated balancing energy results, this control state may provide input for imbalance mechanism simulations (see below).  

\paragraph{Imbalance mechanism implementation} An imbalance clearing method determines the imbalance price. An imbalance settlement method allocates costs and profits among market parties depending on their imbalances. Two design options for market rules are currently implemented for the imbalance mechanism or market: (1) A fixed single price method considers only a default imbalance price for all simulation steps, (2) the imbalance pricing regime used in the Netherlands (see \cite{TenneTTSOB.V.2019}). The latter applies an exogenous conditional probability distribution function to determine the imbalance price. The conditional variables are the DAM price and the balancing control state.  

An overview of implemented design options per market is shown in \ref{apx:agentstrategies}. To implement a new option, a name for the option needs to be defined. Subsequently, conditions for clearing and settlement have to be programmed to the respective market operator methods.

\subsection{Agent process implementation} 
 
\paragraph{Methods of market party agents} During each simulation step, each agent executes a number of methods. Firstly, the agent updates its trade schedule based on the cleared orders from the previous step. Secondly, constraints from redispatch transactions are added to respective assets in its portfolio. Thirdly, the agent runs a portfolio dispatch optimization and determines the available capacity of each asset (see \ref{apx: avcap}). Finally, the agent places orders on various markets. Order placement methods for the DAM, IDM, RDM, and BEM determine timing, quantities, and prices for orders based on the agent's strategies, which are subsequently sent to the respective order books. The imbalance market works without orders because the realized dispatch position and the trading position determine the imbalance volume. However, market party agents manage their imbalance risk by reducing their scheduled imbalances (i.e. expected imbalances which are not yet realized) via trading on the IDM.

\paragraph{Agent strategies} The agents' portfolio dispatch optimization makes use of the optimal power flow functionality of PyPSA. Each agent has a one-node grid model with all its assets implemented as generators, including inter-temporal constraints. The agents' total trade position is implemented as a load in the PyPSA instance. The optimization considers the entire schedules horizon (i.e. all ISPs until the latest traded MTU of the day-ahead market) (for details see \cite{pypsa_docu} and ASAM GitHub repository\cite{AsamGithub}). 
\\
An overview of the currently implemented agent strategies per market is available in \ref{apx:agentstrategies}. These strategies can be extended in the order placement methods of the agent class. However, the detailed strategies and price mark-up determination were outside the scope of this paper.
\begin{figure}[th]
\centering
\includegraphics[width=0.8\textwidth]{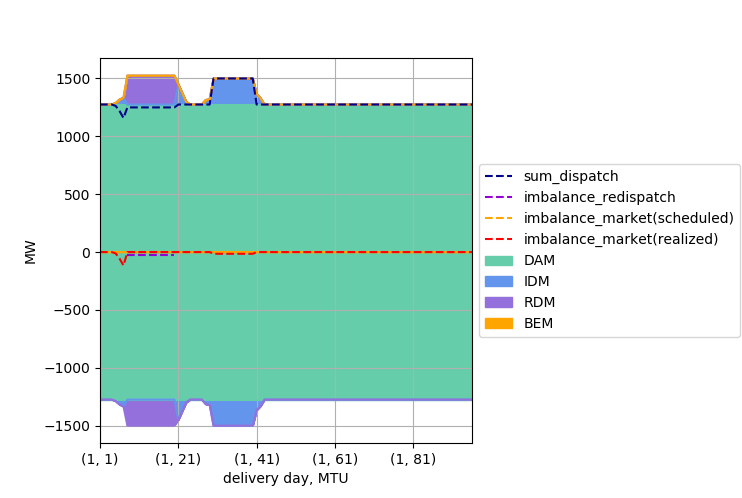}

\caption{Traded volumes on system level after final simulation step (1, 96).}
\label{fig:system overview}
\end{figure}

\paragraph{Methods of grid \& system operator} In each simulation step, the grid \& system operator executes its methods after the market parties have finished their actions. Firstly, new congestions are identified. Secondly, redispatch demand is provided in form of orders to the redispatch market operator. The third action is a market consistency check, followed by an update of the system imbalance.


\section{Simulation results}
\label{sec: results}
This section presents the simulation results of a stylized scenario. The aim of the simulation was to (1) illustrate ASAM outputs, (2) to verify the implementation with a comprehensible scenario, and (3) to show the possible use of functionality and indicators in design studies.
The scenario had two market parties each with two generators. A flat residual load profile was provided exogenously, as well as one period of redispatch and one period of a forecast error (see Annex \ref{apx: scenario inputdata} for input data). As market rules for redispatch resemble the "Reserves for Other Purposes" design in the Netherlands \cite{TennetROP}, which means that all-or-none block orders are allowed and that the market is cleared pay-as-bid (see Annex \ref{apx: scenario inputdata}). For the imbalance mechanism as well, the current Dutch design is used \cite{TenneTTSOB.V.2019} with a conditional probability distribution function based on Dutch data (IBM prices, DAM prices, balancing control state) from 2016 to 2018. We use this stylized scenario for an exemplary research question: How does the all-or-none block order design impact the redispatch performance of the grid operator and the risks and opportunities for market parties?     

Figure \ref{fig:system overview} shows the overall trading situation at the last simulation step. Bought electricity (including downward ancillary services and long imbalance positions) have a positive notation, while sold electricity (including upward ancillary services and short imbalance positions) have a negative sign (see Figure \ref{fig: notation} in Section \ref{sec: glossary} for notation). The figure shows the flat profile of the DAM, one period of RDM transactions, and one period of IDM transactions. In the beginning of the simulation period there is a peak of realized market imbalances. 
\begin{figure}[th]
\centering
\includegraphics[width=1\textwidth]{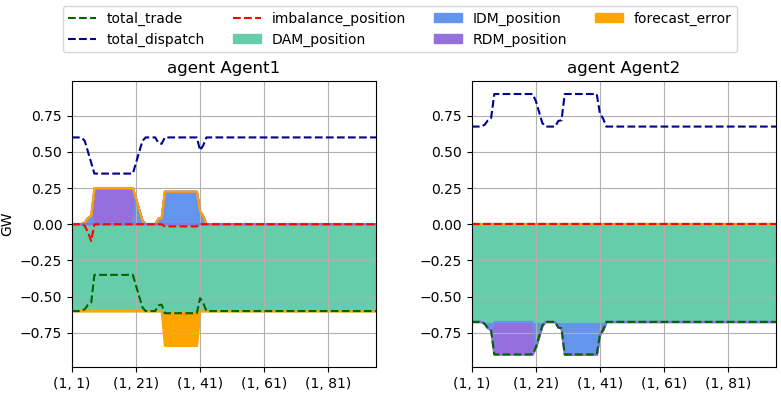}
\caption{Trade position per market party and market after final simulation step (1, 96).}
\label{fig:trade positions}
\end{figure} 

Figure \ref{fig:trade positions} depicts the trade position per market party, including forecast errors. Market party Agent 1 buys nearly the same amount of electricity on the IDM as its short position from forecast errors. As there is only one other market party, it is intuitive that Agent 2 sells this amount on IDM. Furthermore, Agent 1 provides downward redispatch services, and Agent 2 provides upward redispatch to the grid \& system operator. Figure \ref{fig:trade positions} further exhibits imbalances of Agent 1 in advance of the redispatch transaction. As there are a ramp-shaped IDM transactions after the redispatch transactions and only limited IDM volume traded previously, it is likely that the imbalance of Agent 1 stems from ramping limits and that there is not a sufficient amount of time to fully trade the open position on the IDM.          
\begin{figure}[th]
\centering
\includegraphics[width=1\textwidth]{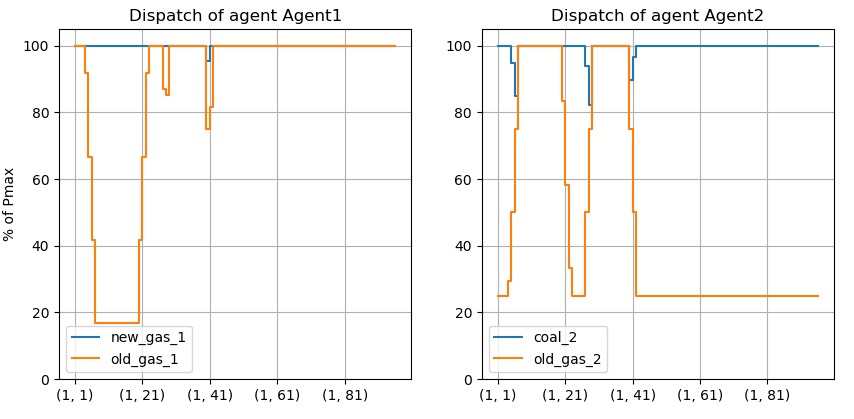}
\caption{Dispatch per asset after final simulation step (1, 96).}
\label{fig:dispatch schedules}
\end{figure} 

Figure \ref{fig:dispatch schedules} shows the dispatch of all generators. Apparently, both generators of Agent 1 are fully dispatched during a majority of market time units. Agent 2 fully dispatches its coal generator and partly commits the old gas generator. This dispatch is as expected because that generator has the highest short-run marginal cost (SRMC) in the system. The downward redispatch is provided by old\_gas\_1, which has the highest SRMC in region north and can therefore provide the best price (i.e. highest buy order) for downward redispatch. In region south only old\_gas\_2 has capacity available for upward redispatch. The stepwise redispatch underlines the suggestion above regarding the imbalance caused by ramping constraints. Moreover, this figure exhibits the need for Agent 1 to buy electricity on IDM during the forecast error period, instead of changing its dispatch: all its generators are fully dispatched, so they cannot cover the short position from forecast errors. Interestingly, Agent 1 reduces its generator dispatch at the beginning and at the end of the forecast error period. This is an effect from generator ramping limitations of Agent 2 while attempting to dispatch the sold electricity during the forecast error period. Both market parties trade additional energy for these ramping periods on the IDM and succeed to avoid imbalances.  
\begin{figure}[th]
\centering
\includegraphics[width=0.8\textwidth]{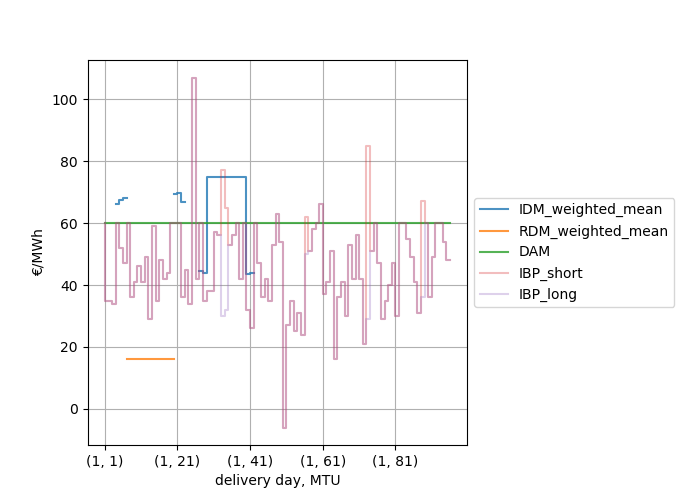}

\caption{Cleared prices.}
\label{fig:cleared prices}
\end{figure} 

Figure \ref{fig:cleared prices} shows the cleared prices of the DAM, IDM, RDM, and IBM\footnote{BEM is not cleared, as explained above.}. Since the residual load has a flat profile, it is intuitive that, simultaneously, the DAM price is steady. Given the agent strategy for DAM order prices (i.e. SRMC bidding), it is further straightforward to observe that the DAM price corresponds to the highest SRMC of the dispatched generators. The weighted mean of the IDM price fluctuates around the DAM price. The price peaks occur during the forecast error period of Agent 1. The RDM weighted mean is actually the price spread of the weighted mean price for upward redispatch and the weighted mean for downward redispatch. The result shows, in line with theory \cite{Hirth2018}, that the grid operator incurs costs (and no profit) to alter the dispatch result of the market parties. Imbalance prices fluctuate, given the exogenous conditional probability distribution of imbalance clearing prices. The imbalance price for short and long positions are most of the time equivalent (i.e. single pricing situation).  
\begin{figure}[th]
\centering
\includegraphics[width=0.8\textwidth]{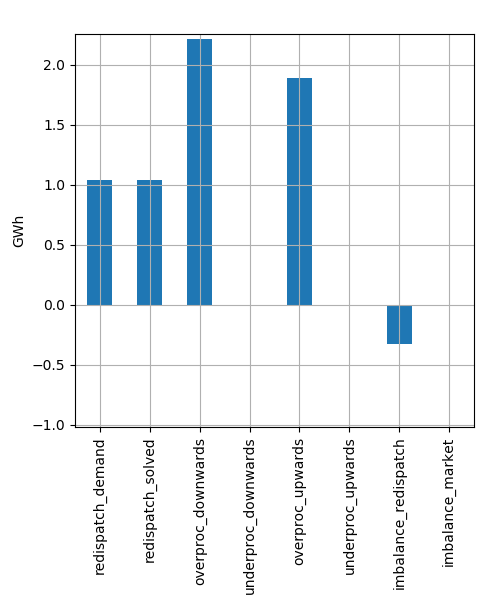}
\caption{Redispatch performance indicators.}
\label{fig: redispatch kpi}
\end{figure} 

Figure \ref{fig: redispatch kpi} displays a summary of the redispatch indicators. The figure shows the result of the simulated market design and the agent strategies. The figure reveals that the entire redispatch demand is solved; however, due to all-or-none orders of the RDM design, the grid operator has procured more redispatch services than needed (i.e. over-procurement). Moreover, the redispatch action for upward and downward are not "energy neutral"\footnote{One could also say: the redispatch action is incomplete.}, thus the grid operator induces imbalances to the system. 
\begin{figure}[th]
\centering
\includegraphics[width=0.8\textwidth]{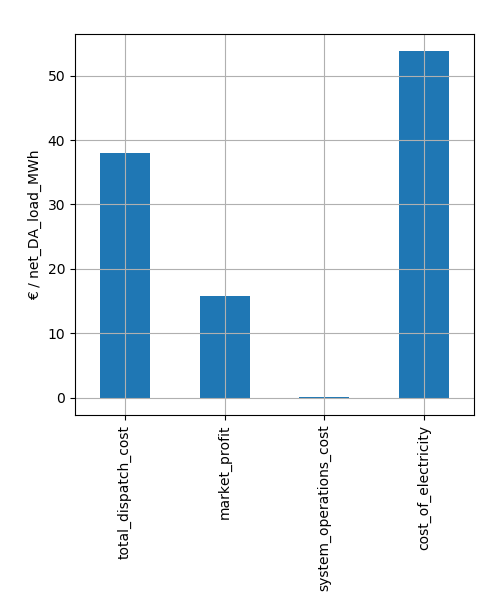}

\caption{Costs and profits on system level.}
\label{fig: profitloss}
\end{figure} 

Figure \ref{fig: profitloss} shows the sum of different costs and profits. The values are normalized to the sum of exogenous day-ahead load (€ per MWh). The system operations cost only reflect the redispatch, as balancing energy is not cleared. In this scenario, system operations cost are only a very small fraction of the cost of electricity. 

\begin{table}[th]
\caption{Prices and quantities of various markets for interdependency analyses.}
\label{tab:interdependency pi}
\footnotesize
\begin{tabular}{@{}lllll@{}}
\toprule
                                                                       & \textbf{BEM} & \textbf{IDM} & \textbf{RDM} & \textbf{DAM} \\ \midrule
prices (median   of w.mean, sell, offered) {[}€\textbackslash{}MWh{]} & 60 & 66 & 68 & 44 \\
prices (median   of  w.mean, buy, offered) {[}€\textbackslash{}MWh{]} & 42 & 37 & 41 &  \\
prices (median   of  w.mean, sell, cleared) {[}€\textbackslash{}MWh{]} &  & 75 & 68 & 60 \\
prices (median   of  w.mean, buy, cleared) {[}€\textbackslash{}MWh{]} &  & 75 & 52 & 60 \\
quantity (sum,   sell, offered) {[}MWh{]} & 1294 & 12625 & 222764 & 30600 \\
quantity (sum,   buy, offered) {[}MWh{]} & 6623 & 52373 & 764194 & 30600 \\
quantity (sum,   sell, cleared) {[}MWh{]} &  & 780 & 731 & 30600 \\
quantity (sum,   buy, cleared) {[}MWh{]} &  & 780 & 813 & 30600 \\
relative   quantity cleared to total quantity cleared {[}\%{]} &  & 2 & 2 & 95 \\
return {[}€{]} &  & 55871 & 91975 & 1836000 \\
relative   return to total return {[}\%{]} &  & 3 & 5 & 93 \\ \bottomrule
\end{tabular}
\end{table}

Table \ref{tab:interdependency pi} shows a number quantity and price indicators per market. When simulating various scenarios and market design, these interdependency indicators can be used to evaluate differences in offered and cleared prices and quantities as well as shifts of return between markets. As each delivery period may be traded per simulation step (i.e. continuous trading), the price indicators reflect the median across the simulations steps of quantity weighted means per delivery period.

\begin{figure}[th]
\centering
\includegraphics[width=1\textwidth]{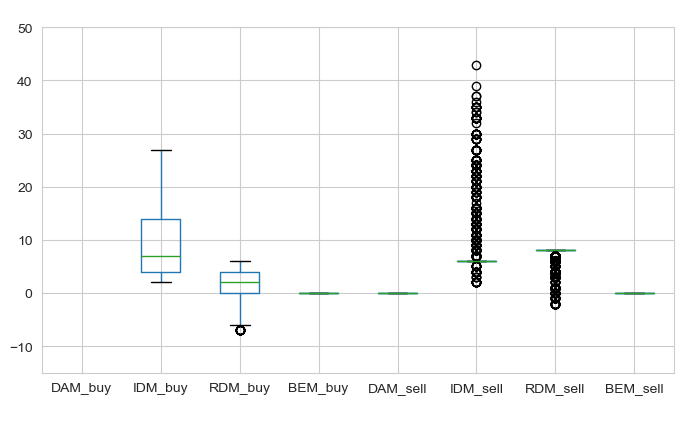}
\caption{Price mark-ups of offered orders. The y-axis is cut at 75 €\textbackslash MWh. Therefore, a few outliers of RDM buy (185 €\textbackslash MWh and 200 €\textbackslash MWh are not visible).}
\label{fig: mark-up}
\end{figure} 

Lastly, all offered or cleared orders may be analyzed regarding mark-ups. Figure \ref{fig: mark-up} exhibits all offered order prices minus the respective SRMC of the associated asset. Negative mark-ups indicate that a offer price becomes "less expensive" (i.e. lower price in case of sell orders, higher price in case of buy orders). When comparing design options, the differences in mark-ups may be visualized with a boxplot.
In this simulation, orders for the DAM\footnote{Buy orders in DAM represent the exogenous residual load and do not have a price.} and BEM do not incorporate price mark-ups, as the agents apply a simple SRMC bidding strategy for these markets. The outliers of the redispatch market (not visible, only described) result from high mark-ups from shut-down costs. Furthermore, the figure shows that the mark-ups for RDM are lower compared to IDM. This is a result of the applied agent-strategies. Although the description of the agent strategies is outside the scope of this paper, it has to be noted that the redispatch mark-ups of the simulation do not include any strategic mark-ups. 

\section{Discussion}
\label{sec: discussion}
The results illustrated the functions and output of ASAM: the model allows for the implementation of various market designs, agent strategies, and detailed representation of market parties. Simulations enable analyses at the agent level (e.g. offers per market and dispatch optimization) and system level (e.g. costs and redispatch performance). Given the high stylization of the simulated scenario, empirical input validation, and calibration, as proposed by \cite{Tesfatsion2018}, was not applicable. However, the simulation of the simple scenario indicated intuitive results regarding transactions, dispatch schedules, prices, and system costs. This model verification\footnote{"[...]the computational model is correctly implemented and working as intended."\cite[ p. 29]{Weidlich2008}} is supplemented by unit tests and integration tests on GitHub \cite{AsamGithub}. 
\\

Section \ref{sec: material and methods} describes the fundamental features of EU market designs, and Section \ref{sec:Aim, scope and main assumptions} explores applied theories, assumptions and simplifications of the model. Based on these sections, model-validity\footnote{"validity of the model relative to the theory" \cite[p. 29]{Weidlich2008}.} shows that, for a European scope, many aspects are represented in large detail in ASAM: portfolio dispatch optimization with inter-temporal asset constraints, bidding process per market, and clearing methods of the DAM, IDM, RDM, and IBM. However, several processes are strongly simplified (e.g. BEM) or not yet implemented (e.g. BCM, coupling of multiple bidding zones). Moreover, physical system models are not yet implemented in or coupled to ASAM (e.g. power flows, system balance, and frequency). Therefore, it has to be noted that the feedback-loop between market results and effects on the physical system remains limited. Furthermore, theory-validation\footnote{"validity of the theory relative to the real-world system" \cite[p. 29]{Weidlich2008}} regarding specific strategic options of agents were outside the scope of this paper.
\\

The exemplary research question regarding the impact of all-or-none block orders on redispatch performance and market parties is supported by ASAM as follows: (1) RDM design is implemented in detail, and thus bidding and clearing of all-or-none block orders is possible. (2) redispatch performance indicators measure under-procurement, over-procurement, and redispatch-induced imbalance to the system. (3) Order analysis and price mark-ups provide insight into risks and opportunities for market parties and the emergent effects on the RDM and other markets. For example, large activated redispatch orders may lead to imbalance risks by ramping constraints. This risk can be mitigated with additional transactions on the IDM (i.e. market interaction). Moreover, start-up and shut-down capacity may be offered for redispatch, without additional risks for the agent of being partially activated in quantity or time (i.e. MTU). Except for additional start-up or shut-down costs, no mark-ups regarding imbalance related to the minimum stable operation or minimum up- or down-time need to be considered. This example thus underpins the suitability of ASAM for its defined target: contributing to the evaluation of various ancillary service acquisition designs, including potential interactions with other acquisition processes and markets.
\\  

Following \cite{Tesfatsion2018} it should be stated for what level of policy advice the model can contribute results. Given the detailed representation of various market processes and the asset portfolios of market parties (as opposed to asset representation only), it can be concluded that ASAM is valid for assessments of policy-readiness levels 4 and 5: \textit{"Policy performance tests using a small-scale model embodying several\footnote{"many" in case of policy-readiness-level 5.} salient real-world aspects"} \cite[p. 17]{Tesfatsion2018}.

\section{Conclusions}
\label{sec: conclusion}
Interactions of markets and ancillary services remain a particular challenge for assessments of policy performance. The literature review showed that, currently, an open-source agent-based model is lacking for design tests of electricity markets and ancillary services in the EU context. The ASAM aims to contribute to this gap by combining the agent-based modeling framework MESA with the power system analyses toolbox PyPSA and providing a basic structure of market processes and performance indicators.
\\
An ASAM simulation of a stylized scenario is used to verify the implementation and validate the model. Furthermore, the scenario addresses an exemplary research question regarding the design of order types in a redispatch market. Thereby, the simulation makes use of the following distinguishing features of ASAM compared to other state-of-the art models:

\begin{itemize}
    \item Configurable generic design variables per market or acquisition process and short-term markets are already implemented (i.e. DAM, IDM, RDM, BEM, IBM) with varying level of detail.
    \item A novel redispatch clearing algorithm, using PyPSA, which allows for complex designs, such as all-or-none orders, limit orders, and multi-MTU block orders.
    \item An intra-day clearing algorithm for continuous trading with open order books.
    \item A day-ahead clearing algorithm, using PyPSA, for sealed auctions, which also allows for complex orders.
    \item Model reporters for indicators regarding market interactions based on offered and cleared orders per market.
    \item A novel policy performance indicator based on price mark-ups, providing insights regarding risks and opportunities for market parties under various ancillary services designs.
    \item Novel redispatch performance indicators, considering over-procurement, under-procurement as well as grid operator caused imbalances.
    \item Strategies of market parties to mitigate imbalance risks via intra-day trading and portfolio dispatch optimization (as opposed to commonly assumed dispatch optimization at the system level). 
\end{itemize}

The simulation results displayed the following implications from order types in redispatch markets with a multi-MTU all-or-none design:
\begin{itemize}
    \item Start-up and shut-down capacity may be offered by market parties for redispatch without additional risk mark-ups for partial order clearing. However, non-strategic mark-ups may include ramping risk and start-up or shut-down costs.
    \item When large all-or-none orders are cleared, market parties may need to ramp before and after the delivery period, as the ramping of generators may be too slow. This would lead to imbalances and additional transactions on IDM for mitigation. This can be considered to be a design-driven market interaction. 
    \item All-or-none orders in a liquid situation lead to over-procurement of redispatch by the grid operator, as orders can not be partially activated. Moreover, it is likely that the grid operator induces imbalances to the system by "incomplete" redispatch activation (i.e. upward and downward redispatch volumes are not equal).  
\end{itemize}
However, the stylized scenario was not suitable for an absolute or relative\footnote{Relative to other design options.}) quantification of the effects above. Quantification would require more real-world aspects.
Nonetheless, the results and discussion show that ASAM is a comprehensive simulation model for the assessment of ancillary services acquisition and market design. The open-source availability of ASAM and its structure (e.g. configurable design variables, combinable with other tools) provide an excellent starting point for researchers conducting robustness tests of policy options.
\\

For further research, it is recommended that ASAM be applied to use-cases to show practicality for ancillary service studies and policy advice. Moreover, it is recommended that ASAM be further developed regarding market processes, coupling with other physical models, and agent behavior. 

\section{Acknowledgements}
The author would like to thank Frank Nobel, Ksenia Poplavskaya and Frank Spaan for the valuable inputs and discussions that contributed to this paper as well as the reviewers for their thoughtful feedback. Moreover, the author would like to thank the open energy modeling community as well as the contributors of PyPSA and MESA for their great work.

\section{Data availability}
The simulation input data, simulation results as well as the simulation model are available on GitHub \cite{AsamGithub}.

\section{Notations, abbreviations, variables}
\label{sec: glossary}

\begin{figure}[th]
\centering
\includegraphics[width=0.8\textwidth]{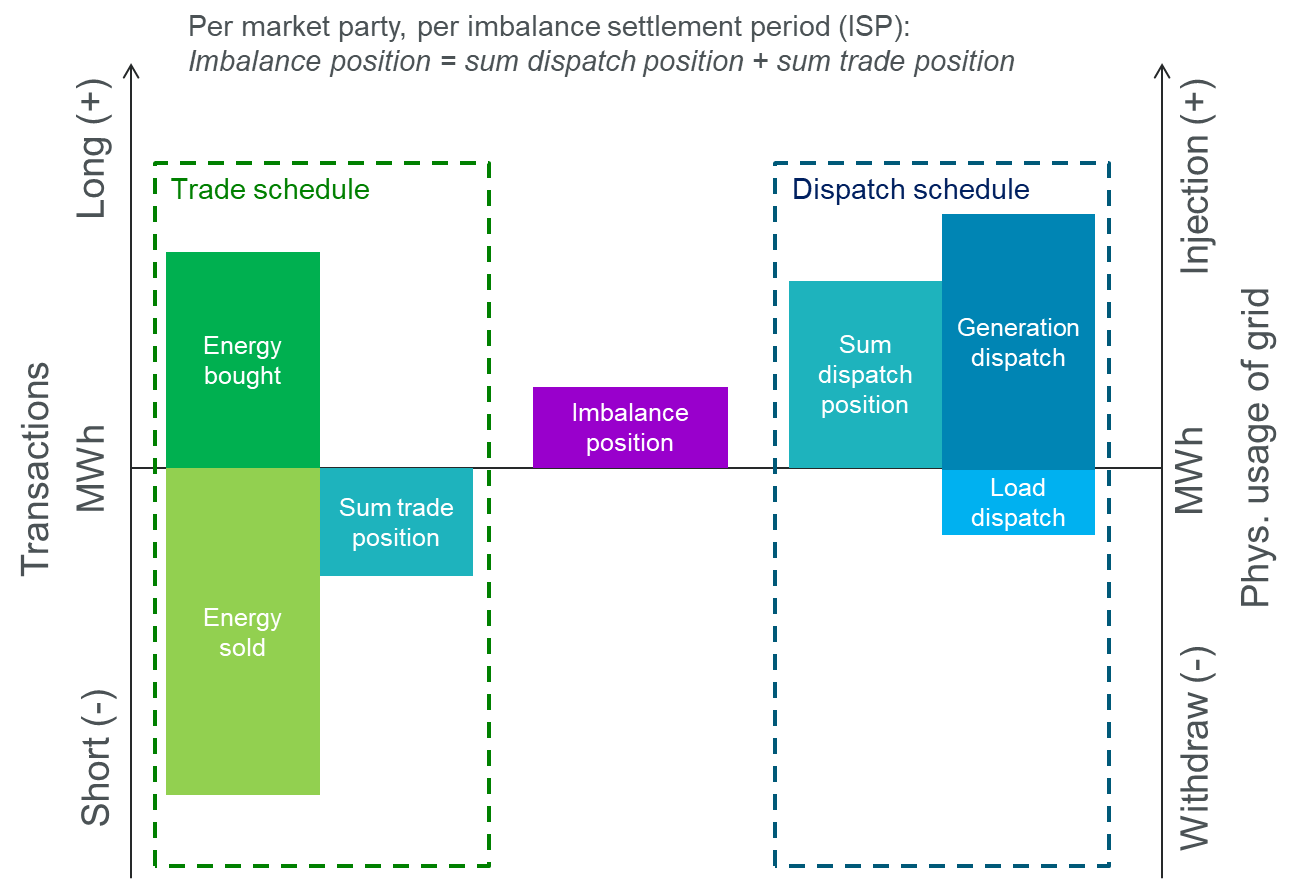}

\caption{Notation of transactions, trade positions, dispatch and physical use of the grid.}
\label{fig: notation}
\end{figure} 

The model uses the notation illustrated in Figure \ref{fig: notation}.

A list of abbreviations is provided in Table \ref{tab:abbrevations}.
\begin{table}
\centering
\caption{List of abbreviations}
\label{tab:abbrevations}
\footnotesize
\begin{tabular}{l l}
\toprule
\textbf{ASAM }& Ancillary Service Acquisition Model\\
\textbf{BCM} & Balancing Capacity market\\
\textbf{BEM} & Balancing Energy Market\\
\textbf{DAM} & Day-ahead Market\\
\textbf{DSO} & Distribution System Operator\\
\textbf{EU} & European Union\\
\textbf{IBM }& Imbalance Market or Mechanism\\
\textbf{ISO }& Independent System Operator\\
\textbf{ISP} & Imbalance Settlement Period\\
\textbf{MTU} & Market Time Unit\\
\textbf{RDM} & Redispatch Market or Mechanism\\
\textbf{SRMC} & Short-run marginal cost\\
\textbf{TSO }& Transmission System Operator\\
\bottomrule
\end{tabular}   
\end{table}

The variables used in the equations of the following sections are listed in table \ref{tab:Variables of markets and agents}. As PyPSA is applied in market clearing methods and in agent dispatch optimization, the variable names are largely aligned with the names of the PyPSA documentation.   
\begin{table}
    \centering
    \caption{Variables of equations regarding markets and agents in ASAM}
    \label{tab:Variables of markets and agents}
    \footnotesize
    \begin{tabular}{l l}
    \toprule
    Variable & Meaning\\
    \midrule
    $t$ & ISP, index for time steps in schedules horizon \\
    ~ &$t\in T=\{0,…|T|-1\}$ \\
    $A$ & Index for agents\\
    $OB$ & Index for order books of different markets\\
    $bu$ & Index for buy order\\
    $se$ & Index for sell order\\
    $n$ & Index for grid areas (respectively agent portfolios)\\
    ~ &$n\in N=\{0,…|N|-1\}$ \\
    $s$ & Index for the generators $s\in S=\{0,…|S|-1\}$\\
    $g_{n,s,t}$ & Dispatch of a generator at a time step [MW].\\
    $u_{n,s,t}$& Time series of binary status variables \\
    ~ &  for inter-temporary constraints $\in\{0,1\}$ \\
    $p\_min\_pu_{n,s,t}$ & Time-variant minimum stable level in p.u. of $P_{nom}$\\
    $p\_max\_pu_{n,s,t}$ & Time-variant maximum output in p.u. of $P_{nom}$\\
    $min\_up\_time_{n,s}$ & Minimum up time [ISP]\\
    $min\_down\_time_{n,s}$ & Minimum down time [ISP]\\
    $srmc_{n,s}$ & Short-run marginal costs\\
    $suc_{n,s,t}$ &Start-up cost when started at t [EUR]\\
    $sdc_{n,s,t}$ &Start-down cost when shut-down at t [EUR]\\
    $ru_{n,s}$& Ramp-up limit [MW/ISP]\\
    $rd_{n,s}$& Ramp-down limit [MW/ISP]\\
    $minSD_{n,s,t}$&Must-run (minimum) scheduled dispatch [MW]\\
    $maxSD_{n,s,t}$&Maximum scheduled dispatch [MW]\\
    $av\_cap\_up_{n,s,t}$&Available upward capacity of asset s[MW]\\
    $av\_cap\_down_{n,s,t}$&Available downward capacity of asset s[MW]\\
    $SD_{n,s,t}$&Scheduled dispatch of asset [MW]\\
    $rr\_av\_cap\_up_{n,s,t}$&Remaining ramp constraint $av\_cap\_up$ [MW]\\
    $rr\_av\_cap\_down_{n,s,t}$&Remaining ramp constraint $av\_cap\_down$ [MW]\\
    $TP_{m,n,t}$&Trade position of agent n at market m [MW]\\
    $FE_{n,t}$&Forecast error of agent n [MW]\\
    $Pmin_{n,s}$&Minimum stable dispatch level [MW]\\
    \bottomrule
    \end{tabular}   
\end{table}

\bibliography{library}
\appendix
\section{Implemented market design options} 
\label{apx:marketrules}
Table \ref{tab:Market rules implemented in ASAM} shows the currently implemented design options per design variable of the markets and acquisition processes. The details can be found on GitHub \cite{AsamGithub}. 
\begin{table}[th]
\caption{Market rules implemented in ASAM}
\label{tab:Market rules implemented in ASAM}
\centering
\scriptsize
\setlength\tabcolsep{0pt} 
\begin{tabular}{>{\hspace{0pt}}p{0.07\linewidth}>{\hspace{0pt}}p{0.173\linewidth}>{\hspace{0pt}}p{0.11\linewidth}>{\hspace{0pt}}p{0.181\linewidth}>{\hspace{0pt}}p{0.095\linewidth}>{\hspace{0pt}}p{0.09\linewidth}>{\hspace{0pt}}p{0.1\linewidth}}
\hline
~ &\textbf{Acquisition\newline method} &\textbf{Pricing\newline method} &\textbf{Order\newline types} & \textbf{Gate-\newline opening\newline time} & \textbf{Gate-\newline closure\newline time} & \textbf{Provider\newline accre-\newline ditation} \\ 
\hline
\textbf{DAM} & single\_hourly\_ auction & uniform & fill\_or\_kill & D-1, MTU 48 & D-1, MTU 49 & all \\ 
~ & exo\_default &  &  &  &  & ~ \\ 
\hline
\textbf{IDM} & continous\_IDM & best\_price & limit\_and\newline\_market & D-1, MTU 56 & MTU-1 & all \\ 
~ & ~ & ~ & limit\_market\newline \_ IDCONS & ~ & ~ & ~ \\ 
\hline
\textbf{RED} & cont\_RED\_thinf & pay\_as\_bid & all\_or\_none\_ISP & D-1, MTU 56 & MTU-1 & all \\ 
~ & cont\_RED\_th0 &  & all\_or\_none\_block &  & MTU-3 & ~ \\ 
~ & cont\_RED\_th5 &  & limit\_ISP &  &  & ~ \\ 
~ & cont\_RED\_th50 &  & limit\_block &  &  & ~ \\ 
~ & ~ &  & IDCONS\_orders &  &  & ~ \\ 
\hline
\textbf{BEM} & control\_states\_ only & none & mFRR  aFRR & D-1, MTU 56 & MTU-1 & all \\ 
\hline
\textbf{IBM} & realtime & Dutch\_IB pricing & na & MTU & MTU & all \\ 
~ & ~ & fixed single price & ~ & ~ & ~ & ~ \\ 
\hline
\textbf{BCM} & ~ & ~ & ~ & ~ & ~ & \\
\hline
\end{tabular}
\end{table}

\section{Implemented agent strategies} 
\label{apx:agentstrategies}
Table \ref{tab: Agent strategies implemented in ASAM} shows the currently implemented agent strategies in ASAM regarding order quantity, order pricing and order timing. The details can be found on GitHub \cite{AsamGithub}. 

\begin{table}[th]
\caption{Agent strategies implemented in ASAM}
\label{tab: Agent strategies implemented in ASAM}
\centering
\small
\resizebox{\linewidth}{!}{
\begin{tabular}{>{\hspace{0pt}}p{0.09\linewidth}>{\hspace{0pt}}p{0.22\linewidth}>{\hspace{0pt}}p{0.33\linewidth}>{\hspace{0pt}}p{0.35\linewidth}} 
\toprule
\textbf{Market} & \textbf{Timing} & \textbf{Quantity} & \textbf{Pricing} \\ 
\midrule
\textbf{DAM} & \newline at\_gate\_closure & \tabitem all & \tabitem srmc \\ 
\hline
\textbf{IDM}& \tabitem instantaneous & \tabitem small\_random\newline
  \tabitem all\_operational\newline
  \tabitem all\_operational\newline
  \_+\_conditional \_start-stop & \tabitem srmc +/-1\newline
  \tabitem marginal\_order\_book\_strategy\newline
  \tabitem marginal\_order\_book\_strategy\newline
  \_+\_startstop\_+partial\_call \\ 
\hline
\textbf{RED}& \tabitem instantaneous & \tabitem small\_random\newline
  \tabitem all\_operational\newline
  \tabitem all\_operational\_+\_start-stop\newline
  \tabitem not\_offered\_+\_start-stop & \tabitem srmc\newline
  \tabitem opportunity mark-up\newline
  \tabitem ramping mark-up\newline
  \tabitem double-score mark-up\newline
  \tabitem start-stop mark-up\newline
  \tabitem partial-call mark\_up\newline
  \tabitem all mark-ups \\ 
\hline
\textbf{BEM} & \tabitem at\_gate\_closure & \tabitem available\_ramp & \tabitem srmc \\ 
\hline
\textbf{IBM} &\tabitem instantaneous\newline
  \tabitem impatience\_curve & \tabitem all\newline
  \tabitem small\_random\newline
  \tabitem impatience\_curve\newline & \tabitem market\_orders\_strategy\newline
  \tabitem marginal\_order\_book\_strategy\newline
  \tabitem impatience\_curve \\
\bottomrule
\end{tabular}
}
\end{table}

\section{Mathematical formulation of redispatch market algorithm}
\label{apx: redisaptch algo}
This appendix describes the basic redispatch algorithm of ASAM mathematically.
Please note the following simplification: As outlined in \cite{Hirth2018}, the order effectivity varies with the order delivery location. This is usually approached with matrixes of so-called power distribution factors. However, since the ASAM redispatch method has no physical electricity grid modeled, it is assumed that all orders from the same area have a similar effectivity. For radial DSO networks, this assumption is fairly appropriate, but for meshed networks, the simplification may affect the accuracy of results. 

In line with the description by \cite{pypsa_docu}, the optimization function can be expressed as follows. Irrelevant variables for this redispatch approach are neglected, such as grid expansion cost, generator investment cost, storage state and capacity, and weighing factors per timestamp. The optimization function minimizes the cost over the schedules horizon.

\begin{equation}
\label{eq: obj redispatch}
 min \left[\sum_{n,s,t}srmc_{n,s}*g_{n,s,t}\right]
\end{equation}
where $n\in N=\{0,…|N|-1\}$ represents the label of the grid areas (i.e. the one-node grids), $t\in T=\{0,…|T|-1\}$ represents the simulated time steps of the schedules horizon, $s\in S=\{0,…|S|-1\}$ is the index for the generators in an area. $srmc_{n,s}$ is the marginal cost and $g_{n,s,t}$ is the generators’ dispatch.

The cost minimization is subject to the following constraints.

\paragraph{Nodal power balance constraint} In every redispatch area (i.e. grid node) the sum of generators dispatch (i.e. cleared redispatch supply orders) must equal the load (i.e. the redispatch demand) $d_{n,t}$.

\begin{equation}
\label{eq: nodal power balance}
 \sum_{n,s,t}g_{n,s,t}=\sum_{n,t} d_{n,t} ~ \forall ~n,s,t
\end{equation}
Operating-level constraint. All generators may only be dispatched within their operating level from $Pmin$ to $Pnom$. As these two parameters may vary over time, PyPSA expresses this generator availability in per unit of the nominal generator capacity $Pnom$ for all timestamps $t$. 
\begin{equation}
\label{eq: binary variables}
u_{n,s,t}* p\_min\_pu_{n,s,t} *Pnom_{n,s} \leq g_{n,s,t} \leq u_{n,s,t}* p\_max\_pu_{n,s,t}* Pnom_{n,s}
\end{equation}
 where $u_{n,s,t}\in\{0,1\}$ is a time series of binary status variables used to implement inter-temporary constraints. 

\paragraph{Minimum up-time constraint} Orders with a delivery duration larger than 1 MTU (i.e. block orders) make use of the minimum up-time constraint. Let $min\_up\_time \in \mathbb{N}$ be the number of MTUs which a generator at least has to operate when being dispatched. 

\begin{equation}
\label{eq: min uptime}
\sum_{t'=t}^{t+min\_up\_time} u_{n,s,t'}\leq min\_up\_time(u_{u,s,t}-u_{n,s,t-1})~ \forall~ n,s,t
\end{equation}
This implies that when a generator is started in $t$, then $u_{n,s,t-1}=0$, $u_{n,s,t}=1$ and $u_{n,s,t}-u_{n,s,t-1}=1$. Consequently, the generator has to run for at least $min\_up\_time$ periods.

\paragraph{Additional block order constraint} Generators representing block orders must be dispatched with the same quantity for all MTUs of the order duration (i.e. $min\_up\_time$). This is not foreseen in a usual PyPSA power flow. However, PyPSA allows for formulating custom constraints. The block order constraint can be expressed as follows:
\begin{equation}
g_{n,s,t}=g_{n,s,t-1}~\forall~\{t\mid~p\_max\_pu_{n,s,t-1 } = 1 \land t\in DeliveryPeriod \}
\end{equation}

This constraint thus applies to the delivery period of the order. 

\paragraph{Equilibrium constraint with a threshold} The equilibrium constraint requires that the sum of cleared order quantity in upward direction $g_{up,s,t}$ equals the sum of cleared order quantity in downward direction $g_{down,s,t}$, while taking into account a threshold.

\begin{equation}
-threshold \leq \sum_s g_{up,s,t} - \sum_s g_{down,s,t} \leq threshold ~\forall t
\end{equation}

\section {Available capacity determination per asset}
\label{apx: avcap}

The portfolio dispatch optimization results in a scheduled dispatch for each asset of an agent ($SD\_{n,s,t}$). Subsequently, the available capacity of each asset ($av\_cap\_up_{n,s,t}$, $av\_cap\_down_{n,s,t}$) is calculated as well as the available capacity under consideration of the remaining ramp ($rr\_av\_cap\_up_{n,s,t}$, $rr\_av\_cap\_down_{n,s,t}$). For the latter, the scheduled ramps before and after MTU $t$ are deducted from the ramp limits of the asset.
\begin{equation}
av\_cap\_up_{n,s,t}=p\_max\_pu{(n,s,t}*Pnom_{n,s,t}-SD_{n,s,t}
\end{equation}

\begin{equation}
av\_cap\_down_{n,s,t}=SD_{n,s,t}  -p\_min\_pu_{n,s,t}*Pnom_{n,s,t}
\end{equation}
\begin{multline}
rr\_av\_cap\_up_{n,s,t}=\\max\left(min\left(
\begin{array}{l}
ru_{n,s}*Pnom_{n,s}-(SD_{n,s,t}-SD_{n,s,t-1}),\\
ru_{n,s}*Pnom_{n,s}-(SD_{n,s,t+1}-SD_{n,s,t} ),\\
av\_cap\_up_{n,s,t}
\end{array}\right),0\right)
\end{multline}
\\
\\
\begin{multline}
rr\_av\_cap\_down_{n,s,t}=\\max\left(min\left(
\begin{array}{l}
rd_{n,s}*Pnom_{n,s}-(SD_{n,s,t-1}-SD_{n,s,t}),\\
rd_{n,s}*Pnom_{n,s}-(SD_{n,s,t}-SD_{n,s,t+1} ),\\
av\_cap\_down_{n,s,t}
\end{array}\right),0\right)
\end{multline}
\section{Input data of the simulation scenario}
\label{apx: scenario inputdata}

\begin{table}[th]
\caption{Simulation setting}
\label{tab:simtask}
\footnotesize
\begin{tabular}{@{}ll@{}}
\toprule
\textbf{parameter}        & \textbf{value}                               \\ \midrule
start\_day                & 1                                            \\
start\_MTU                & 1                                            \\
number\_steps             & 96                                           \\
seed                      & 3                                            \\
forecast\_errors          & exogenous                                   \\
congestions               & exogenous                                   \\
residual\_load\_scenario  & flat profile 0.85 p.u of generator capacity. \\
Day-ahead\_load\_scenario & flat profile 0.95 p.u of generator capacity. \\ \bottomrule
\end{tabular}
\end{table}

\begin{table}[th]
\caption{Exogenous forecast errors and congestions}
\label{tab:FE and Congestion}
\footnotesize
\begin{tabular}{@{}lll@{}}
\toprule
\textbf{parameter}   & \textbf{forecast\_errors} & \textbf{congestions} \\ \midrule
identification\_day  & 1                         & 1                    \\
identification\_MTU  & 2                         & 2                    \\
start\_day           & 1                         & 1                    \\
start\_time          & 30                        & 8                    \\
end\_day             & 1                         & 1                    \\
end\_time            & 40                        & 20                   \\
error\_magnitude\_pu & -0.4                      &                      \\
error\_of\_agents    & Agent1                    &                      \\
redispatch\_quantity   &                           & 80                   \\
down\_area           &                           & North                \\
up\_area             &                           & South                \\ \bottomrule
\end{tabular}
\end{table}

\begin{table}[th]
\caption{Asset portfolios}
\label{tab:asset portfolios}
\footnotesize
\begin{tabular}{@{}lllll@{}}
\toprule
\textbf{asset\_name}    & \textbf{old\_gas\_1} & \textbf{new\_gas\_1} & \textbf{old\_gas\_2} & \textbf{coal\_2} \\ \midrule
asset\_owner            & Agent1               & Agent1               & Agent2               & Agent2           \\
Type                    & CCGT old 1           & CCGT new             & CCGT old 2           & coal new         \\
pmax                    & 300                  & 300                  & 300                  & 600              \\
pmin                    & 50                   & 20                   & 50                   & 400              \\
location                & North                & North                & South                & South            \\
srmc                    & 50                   & 40                   & 60                   & 35               \\
ramp\_limit\_up         & 0.25                 & 0.25                 & 0.25                 & 0.25             \\
ramp\_limit\_down       & 0.25                 & 0.25                 & 0.25                 & 0.25             \\
ramp\_limit\_start\_up  & 0.25                 & 0.25                 & 0.25                 & 0.25             \\
ramp\_limit\_shut\_down & 0.25                 & 0.25                 & 0.25                 & 0.25             \\
min\_down\_time         & 12                   & 8                    & 12                   & 20               \\
min\_up\_time           & 12                   & 8                    & 12                   & 20               \\
start\_up\_cost         & 15000                & 6000                 & 15000                & 50000            \\
shut\_down\_cost        & 15000                & 6000                 & 15000                & 50000            \\ \bottomrule
\end{tabular}
\end{table}

\begin{table}[th]
\caption{Market rules setting}
\label{tab:Market rules setting}
\setlength\tabcolsep{0pt}
\scriptsize
\begin{tabular}{@{}llllll@{}}
\toprule
\textbf{}               & \textbf{DAM}            & \textbf{IDM}       & \textbf{RDM}         & \textbf{BEM}          & \textbf{IBM}       \\ \midrule
gate\_opening\_time     & D-1, MTU 44             & D-1, MTU 56        & D-1, MTU 56          & D-1, MTU 56           & MTU        \\
gate\_closure\_time     & D-1, MTU 45             & MTU-1      & MTU-3        & MTU-2         & deliveryMTU        \\
acquisition\_method     & single\_hourly\newline \_auction & continuous          & cont\_RDM\_thinf     & control\_states\_only & realtime           \\
pricing\_method         & uniform                 & best\_price        & pay\_as\_bid         &                       & Dutch\_IB\_pricing \\
order\_types            & fill\_or\_kill          & limit\_and\newline \_market & all\_or\_none\newline \_block & FRR                   & na                 \\
provider\_accreditation & all                     & all                & all                  & all                   & all                \\\bottomrule
\end{tabular}
\end{table}

\begin{table}[th]
\caption{Agent strategies per market and ancillary services process}
\label{tab:agent strategy setting}
\footnotesize
\begin{tabular}{@{}ll@{}}
\toprule
parameter           & value                         \\ \midrule
agent               & All                           \\
DAM\_quantity         & all                           \\
IDM\_quantity         & random                        \\
RDM\_quantity         & all\_plus\_startstop          \\
DAM\_pricing        & srmc                          \\
IDM\_pricing        & marginal\_orderbook\_strategy \\
RDM\_pricing        & all\_markup                   \\
IDM\_timing         & instant                       \\
RDM\_timing         & instant                       \\
IBM\_quantity         & impatience\_curve             \\
IBM\_pricing        & impatience\_curve             \\
IBM\_timing         & impatience\_curve             \\
BEM\_quantity         & available\_ramp               \\
BEM\_pricing        & srmc                          \\
BEM\_timing         & at\_gate\_closure             \\
ramp\_limits        & True                          \\
start\_stop\_costs  & True                          \\
min\_up\_down\_time & True                          \\ \bottomrule
\end{tabular}
\end{table}

\end{document}